# Steganalysis of Transcoding Steganography

Artur Janicki, Wojciech Mazurczyk, Krzysztof Szczypiorski
Warsaw University of Technology, Institute of Telecommunications
Warsaw, Poland, 00-665, Nowowiejska 15/19
e-mail: {ajanicki, wmazurczyk, ksz}@tele.pw.edu.pl

**Abstract**. TranSteg (Trancoding Steganography) is a fairly new IP telephony steganographic method that functions by compressing overt (voice) data to make space for the steganogram by means of transcoding. It offers high steganographic bandwidth, retains good voice quality and is generally harder to detect than other existing VoIP steganographic methods. In TranSteg, after the steganogram reaches the receiver, the hidden information is extracted and the speech data is practically restored to what was originally sent. This is a huge advantage compared with other existing VoIP steganographic methods, where the hidden data can be extracted and removed but the original data cannot be restored because it was previously erased due to a hidden data insertion process. In this paper we address the issue of steganalysis of TranSteg. Various TranSteg scenarios and possibilities of warden(s) localization are analyzed with regards to the TranSteg detection. A steganalysis method based on MFCC (Mel-Frequency Cepstral Coefficients) parameters and GMMs (Gaussian Mixture Models) was developed and tested for various overt/covert codec pairs in a single warden scenario with double transcoding. The proposed method allowed for efficient detection of some codec pairs (e.g., G.711/G.729), whilst some others remained more resistant to detection (e.g., iLBC/AMR).

## 1. Introduction

TranSteg (Transcoding Steganography) is a new steganographic method that has been introduced recently by Mazurczyk et al. [21]. It is intended for a broad class of multimedia and real-time applications, but its main foreseen application is IP telephony. TranSteg can also be exploited in other applications and services (like video streaming) or wherever a possibility exists to efficiently compress the overt data (in a lossy or lossless manner).

TranSteg, like every steganographic method, can be described by the following set of characteristics: its steganographic bandwidth, its undetectability, and the steganographic cost. The term "steganographic bandwidth" refers to the amount of secret data that can be sent per time unit when using a particular method. Undetectability is defined as the inability to detect a steganogram within a certain carrier. The most popular way to detect a steganogram is to analyze the statistical properties of the captured data and compare them with the typical values for that carrier. Lastly, the steganographic cost characterizes the degradation of the carrier caused by the application of the steganographic method. In the case of TranSteg, this cost can be expressed by providing a measure of the conversation quality degradation induced by transcoding and the introduction of an additional delay.

The general idea behind TranSteg is as follows (Fig. 1). RTP [28] packets carrying the user's voice are inspected and the codec originally used for speech encoding (here called the overt codec) is determined by analyzing the PT (Payload Type) field in the RTP header (Fig. 1.1). If typical transcoding occurs then the original voice frames are usually recoded using a different speech codec to achieve a smaller voice frame (Fig. 1.2). But in TranSteg an appropriate covert codec for the overt one is selected. The application of the covert codec yields a comparable voice quality but a smaller voice payload size than originally. Next, the voice stream is transcoded, but the original, larger voice payload size and the codec type indicator are preserved (the PT field is left unchanged). Instead, after placing the transcoded voice of a smaller size inside the original payload field, the remaining free space is filled with hidden data (Fig. 1.3). Of course, the steganogram does not necessarily need to be inserted at the end of the payload field. It can be spread across this field or mixed with voice data as well. We assume that for the purposes of this paper it is not crucial which steganogram spreading mechanism is used, and thus it is out of the scope of this work.

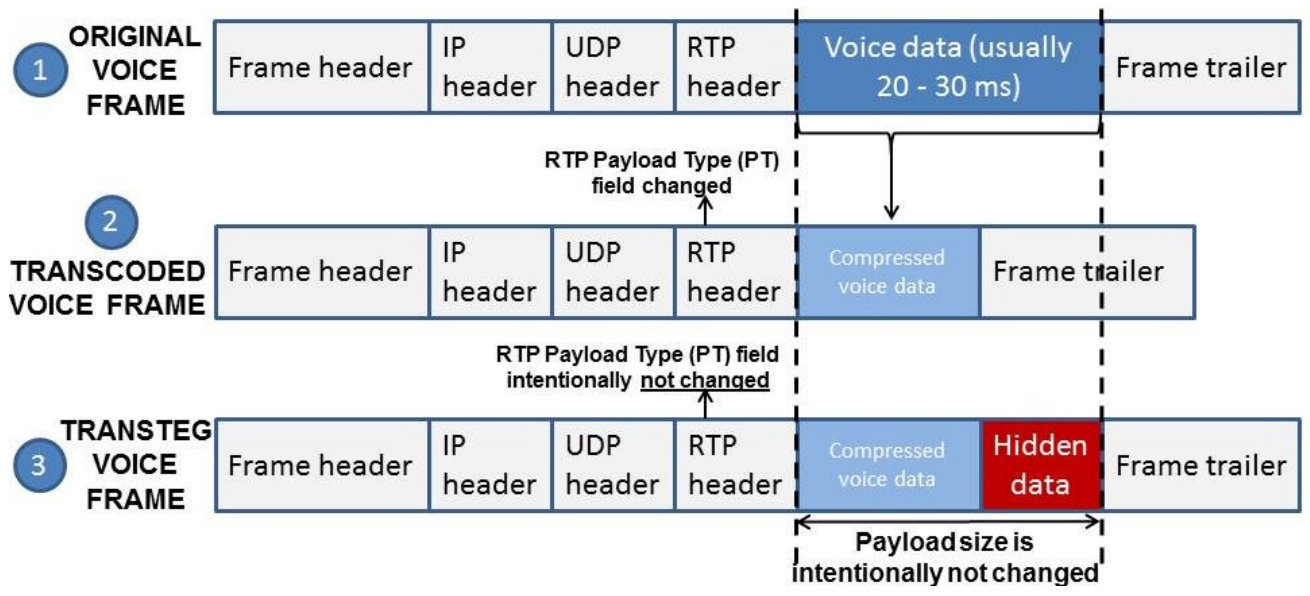

**Fig. 1:** Frame bearing voice payload encoded with overt codec (1), typically transcoded (2), and encoded with covert codec (3)

The performance of TranSteg depends, most notably, on the characteristics of the pair of codecs: the overt codec originally used to encode user speech and the covert codec utilized for transcoding. In ideal conditions the covert codec should not significantly degrade user voice quality compared to the quality of the overt codec (in an ideal situation there should be no negative influence at all). Moreover, it should provide the smallest achievable voice payload size, as this results in the most free space in an RTP packet to convey a steganogram. On the other hand, the overt codec in an ideal situation should result in the largest possible voice payload size to provide, together with the covert codec, the highest achievable steganographic bandwidth. Additionally, it should be commonly used, to avoid arousing suspicion.

In [21] a proof of concept implementation of TranSteg was subjected to experimental evaluation to verify whether it is feasible. The obtained experimental results proved that it offers a high steganographic bandwidth (up to 32 kbit/s for G.711 as overt and G.726 as covert codecs) while introducing delays of about 1 ms and still retaining good voice quality.

In [13] the authors focused on analyzing how the selection of speech codecs affects hidden transmission performance, that is, which codecs would be the most advantageous ones for TranSteg. The results made it possible to recommend 10 pairs of overt/covert codecs which can be used effectively in various conditions depending on the required steganographic bandwidth, the allowed steganographic cost, and the codec used in the overt transmission. In particular, these pairs were grouped into three classes based on the steganographic cost they introduced (Fig. 2). The pair G.711/G.711.0 is costless; nevertheless, it offers a remarkably high steganographic bandwidth, on average more than 31 kbps. However, caution must be taken, as the G.711.0 bitrate is variable and depends on an actual signal being transmitted in the overt channel. Also the AMR (Adaptive Multi-Rate) codec working in 12.2 kbps mode proved to be very efficient as the covert codec for TranSteg.

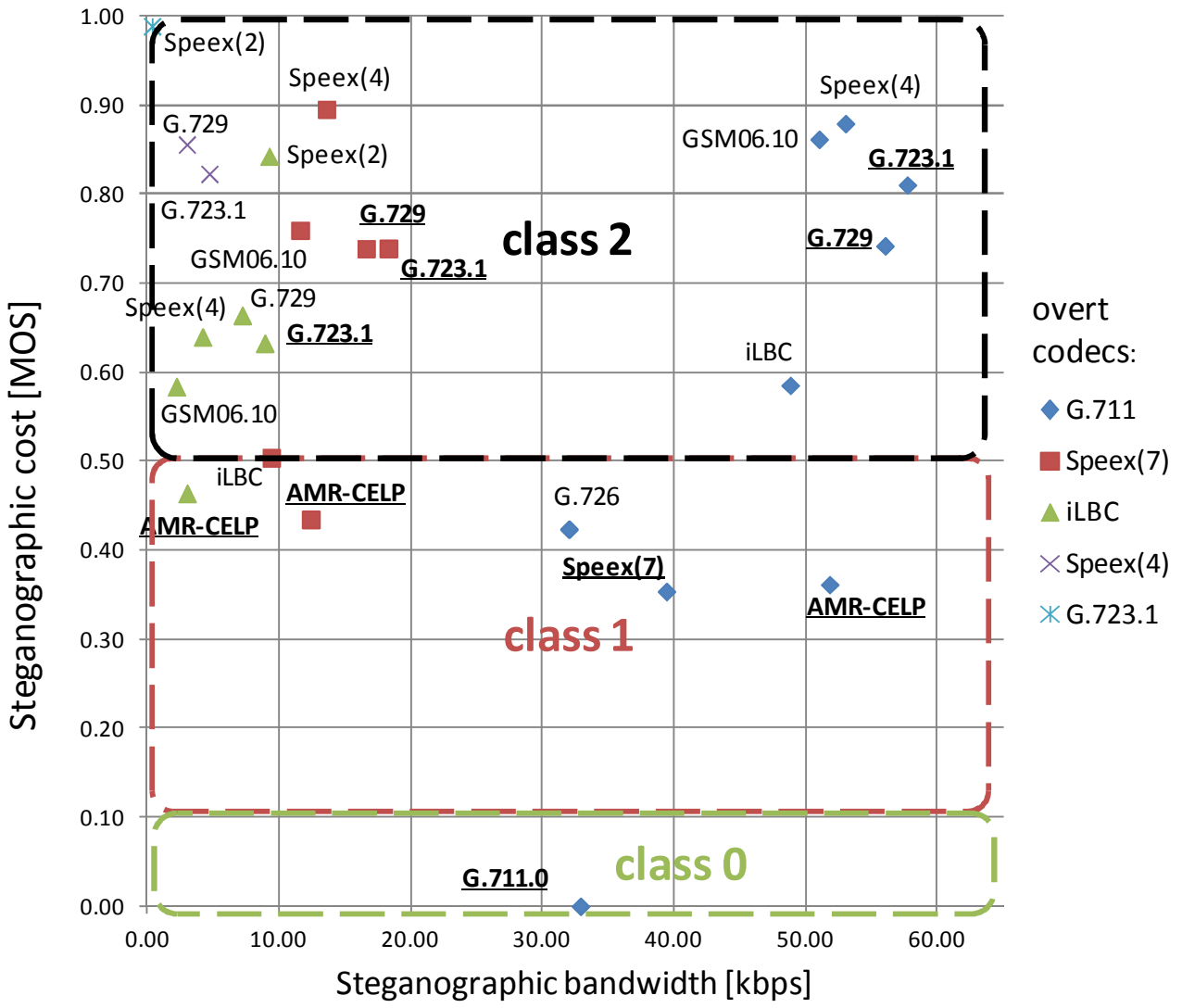

**Fig. 2**: Steganographic cost against the steganographic bandwidth for the tested overt/covert codec pairs. The labels inform about the covert codec [13]

In this paper our main contribution is to develop an effective steganalysis method for TranSteg on the assumption that we are able to capture and analyze only the voice signal near the receiver. We want to verify whether, based only on analysis of this signal, it is possible to detect TranSteg utilization for different voice codecs applied (both overt and covert).

The rest of the paper is structured as follows. Section 2 presents related work on IP telephony steganalysis. Section 3 describes various hidden communication scenarios for TranSteg and discusses its detection possibilities considering various locations of warden(s). Section 4 presents the experimental methodology and results obtained. Finally, Section 5 concludes our work.

## 2. Related Work

Many steganalysis methods have been proposed so far. However, specific VoIP steganography detection methods are not so widespread. In this section we consider only these detection methods that have been evaluated and proved feasible for VoIP. It must be emphasized that many so-called audio steganalysis methods were also developed for detection of hidden data in audio files (so called audio steganography). However, they are beyond the scope of this paper.

Statistical steganalysis for LSB (Least Significant Bits) based VoIP steganography was proposed by Dittmann et al. [5]. They proved that it was possible to detect hidden communication with almost a 99% success rate on the assumption that there are no packet losses and the steganogram is unencrypted/uncompressed.

Takahasi and Lee [29] described a detection method based on calculating the distances between each audio signal and its de-noised residual by using different audio quality metrics. Then a Support Vector Machine (SVM) classifier is utilized for detection of the existence of hidden data. This scheme was tested on LSB, DSSS (Direct Sequence Spread Spectrum), FHSS (Frequency-Hopping Spread Spectrum) and Echo hiding methods and the results obtained show that for the first three algorithms the detection rate was about 94% and for the last it was about 73%.

A Mel-Cepstrum-based detection, known from speaker and speech recognition, was introduced by Kraetzer and Dittmann [18] for the purpose of VoIP steganalysis. On the assumption that a steganographic message is not permanently embedded from the start to the end of the conversation, the authors demonstrated that detection of an LSB-based steganography is efficient with a success rate of 100%. This work was further extended by [19] employing an SVM classifier. In [17] it was shown for an example of VoIP steganalysis that channel character specific detection performs better than when channel characteristic features are not considered.

Steganalysis of LSB steganography based on a sliding window mechanism and an improved variant of the previously known Regular Singular (RS) algorithm was proposed by Huang et al. [12]. Their approach provides a 64% decrease in the detection time over the classic RS, which makes it suitable for VoIP. Moreover, experimental results prove that this solution is able to detect up to five simultaneous VoIP covert channels with a 100% success rate.

Huang et al. [11] also introduced the steganalysis method for compressed VoIP speech that is based on second statistics. In order to estimate the length of the hidden message, the authors proposed to embed hidden data into sampled speech at a fixed embedding rate, followed by embedding other information at a different level of data embedding. Experimental results showed that this solution makes it possible not only to detect hidden data embedded in a compressed VoIP call, but also to accurately estimate its size.

Steganalysis that relies on the classification of RTP packets (as steganographic or non-steganographic ones) and utilizes specialized random projection matrices that take advantage of prior knowledge about the normal traffic structure was proposed by Garateguy et al. [8]. Their approach is based on the assumption that normal traffic packets belong to a subspace of a smaller dimension (first method), or that they can be included in a convex set (second method). Experimental results showed that the subspace-based model proved to be very simple and yielded very good performance, while the convex set-based one was more powerful, but more time-consuming.

Arackaparambil et al. [1] analyzed how, in distribution-based steganalysis, the length of the window of the detection threshold and in which the distribution is measured, should be depicted to provide the greatest chance of success. The results obtained showed how these two parameters should be set for achieving a high rate of detection, whilst maintaining a low rate of false positives. This approach was evaluated based on real-life VoIP traces and a prototype implementation of a simple steganographic method.

A method for detecting CNV-QIM (Complementary Neighbor Vertices-Quantisation Index Modulation) steganography in G.723.1 voice streams was described by Li and Huang [20]. This approach is to build the two models, a distribution histogram and a state transition model, to quantify the codeword distribution characteristics. Based on these two models, feature vectors for training the classifiers for steganalysis are obtained. The technique is implemented by constructing an SVM classifier and the results show that it can achieve an average detection success rate of 96% when the duration of the G.723.1 compressed speech bit stream is less than 5 seconds.

In this paper we develop a TranSteg steganalysis method based on the Mel-Frequency Cepstral Coefficients (MFCC) and Gaussian Mixture Models (GMMs). This method will be applied to various overt/covert codec configurations in the TranSteg technique and its effectiveness will be verified.

## 3. TranSteg Detection Possibilities

It must be emphasized that currently for network steganography, as well as for digital media (image, audio, video files) steganography, there is still no universal "one size fits all" detection solution, so steganalysis methods must be adjusted precisely to the specific information-hiding technique (see Section 2).

Typically it is assumed that the detection of hidden data exchange is left for the warden [6]. In particular it:

- is aware that users can be utilizing hidden communication to exchange data in a covert manner;
- has a knowledge of all existing steganographic methods, but not of the one used by those users;
- is able to try to detect, and/or interrupt the hidden communication.

Let us consider the possible hidden communication scenarios (S1-S4 in Fig. 2), as they greatly influence the detection possibilities for the warden. For VoIP steganography, there are three possible localizations for a warden (denoted in Fig. 2 as W1-W3). A node that performs steganalysis can be placed near the sender, or receiver of the overt communication or at some intermediate node. Moreover, the warden can monitor network traffic in single (centralized warden) or multiple locations (distributed warden). In general, the localization and number of locations in which the warden is able to inspect traffic influences the effectiveness of the detection method.

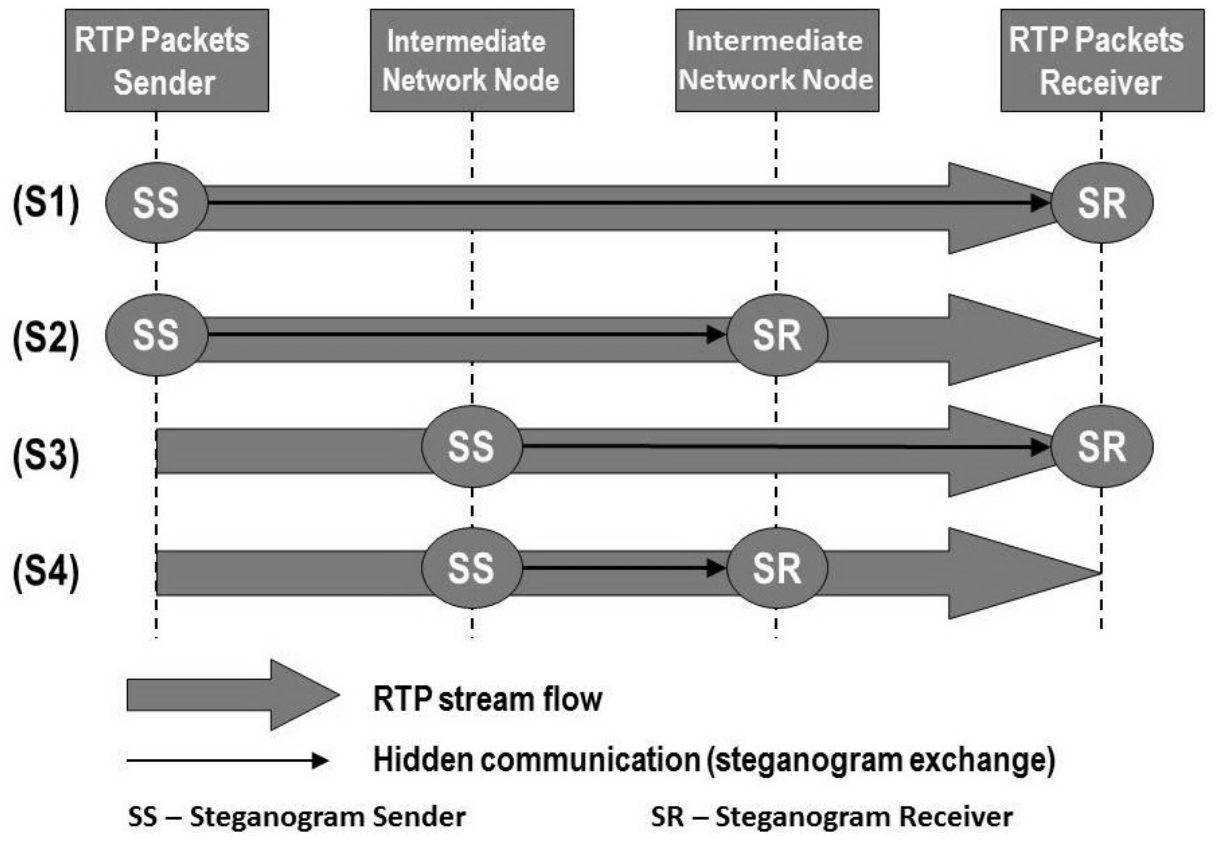

**Fig. 2:** Hidden communication scenarios for VoIP

For TranSteg-based hidden communication, we assume that the warden will not be able to "physically listen" to the speech carried in RTP packets because of the privacy issues related with this matter. This means that the warden will be capable of capturing and analyzing the payload of each RTP packet, but not capable of replaying the call's conversation (its content).

It is worth noting that communication via TranSteg can be thwarted by certain actions undertaken by the wardens. The method can be defeated by applying random transcoding to every non-encrypted VoIP connection to which the warden has access. Alternatively, only suspicious connections may be subject to transcoding. However, such an approach would lead to a deterioration of the quality of conversations. It must be emphasized that not only steganographic calls would be affected – the non-steganographic calls could also be “punished”.

To summarize, the successful detection of TranSteg mainly depends on:

- the location(s) at which the warden is able to monitor the modified RTP stream;
- the utilized TranSteg scenario (S1-S4);
- the choice of the covert and overt codec;
- whether encryption of RTP streams is used.

Let us now consider the distributed warden. When it inspects traffic in at least two localizations, three cases are possible:

- **DWC1:** When the warden inspects traffic in localizations in which RTP packet payloads are coded with overt and then with covert codec (e.g. in scenario S2 localizations W2&W3; in S3 localizations W1&W2). In that case, simple comparison of payloads of certain RTP packets is enough to detect TranSteg.
- **DWC2:** When the warden inspects traffic in localizations in which there is no change of transcoded traffic (e.g. scenario S1 and any two localizations; S2 and localizations W1&W2). In that case, comparing payloads of certain RTP packets is useless as they are exactly the same. However, other detection techniques may be applied here. First, packets can undergo a codec validity test, i.e., they can be checked to determine if selected fields of their payload correspond to the codec type declared in the RTP header. This method can lead to successful detection of TranSteg in most cases. For example, in TranSteg with the Speex as the overt and G.723.1 as the covert codecs pair, if Speex is expected then the first 5 bits of the payload are supposed to contain the wideband flag and the mode type, while the first 6 bits of the G.723.1 payload contain one of the prediction coefficients, so they are variable. Another method consists of simply trying to decode speech with a codec declared in the RTP header. The output signal usually must not be exposed to any human due to the privacy issues mentioned earlier; however, it can undergo voice activity detection (VAD) to check if it contains a speech-like signal [25]. However, it must be noted that if encryption of the data stream is applied e.g. by means of the most popular SRTP (Secure RTP) [2] protocol, then the abovementioned techniques would most likely fail.
- **DWC3:** When the warden inspects traffic in localizations in which the voice is coded with overt codec (scenario S4 and localizations W1&W3). In that case, only if lossless TranSteg transcoding was utilized (e.g. for G.711 as overt and G.711.0 as covert codecs) then the payload values are the same and TranSteg detection is impossible. For other overt/covert codecs pairs, comparison of payloads of certain RTP packets would be enough to detect TranSteg.

If the warden is capable of inspecting traffic solely in a single localization (the more realistic assumption), then the detection is harder to accomplish than for a distributed warden. Also three cases are possible:

- **SLWC1:** The warden analyzes the traffic that has not yet been subjected to transcoding caused by TranSteg and the voice is coded with overt codec (scenarios S3 and S4, localization W1). In that case, it is obvious that TranSteg detection is impossible.
- **SLWC2:** The warden analyzes the traffic that has been subjected to TranSteg transcoding and the voice is coded with covert codec (e.g. scenario S1 and any localization; S2 and localization W1 or W2). This situation is the same as for case DWC2 for a distributed warden.
- **SLWC3:** The warden analyzes the traffic that has been subjected to TranSteg re-transcoding and the voice is again coded with overt codec (scenarios S2 and S4, localization W3). This situation is similar to the case DWC3 for a distributed warden, if lossless TranSteg transcoding was utilized. If a pair of lossy overt/covert codecs is used, the detection is not trivial as only re-transcoded, but encoded with an overt codec voice signal is available.

Table I summarizes the abovementioned TranSteg detection possibilities. It must be emphasized that if encryption of RTP streams is performed, then for scenarios S1-S3 it further masks TranSteg utilization and defeats the simple steganalysis methods indicated below. For scenario S4, encryption prevents TranSteg usage.

Table I Comparison of TranSteg detection possibilities

<table>
<tr><th>Case</th><th>Voice encoded with</th><th>Scenarios/Localizations</th><th>Steganalysis method</th></tr>
<tr><td>DWC1</td><td>Overt-covert</td><td>S2 / W2&W3<br>S3 / W1&W2<br>S4 / W1&W2</td><td>RTP payload comparison</td></tr>
<tr><td>DWC2</td><td>Covert</td><td>S1 / W1&W2 or W2&W3 or W1&W3<br>S2 / W1&W2<br>S3 / W2&W3</td><td>Codec validity test, VAD</td></tr>
<tr><td rowspan="2">DWC3</td><td rowspan="2">Overt<br>(at transmitter & re-transcoded)</td><td rowspan="2">S4 / W1&W3</td><td>For lossless TranSteg transcoding:<br>impossible to detect</td></tr>
<tr><td>For lossy TranSteg transcoding:<br>RTP payload comparison</td></tr>
<tr><td>SLWC1</td><td>Overt codec<br>(at transmitter)</td><td>S3, S4 / W1</td><td>RTP payload comparison</td></tr>
<tr><td>SLWC2</td><td>Covert codec</td><td>S1 / W1 or W2 or W3<br>S2 / W1 or W2<br>S3 / W2 or W3<br>S4 / W2</td><td>Codec validity test, VAD</td></tr>
<tr><td rowspan="2">SLWC3</td><td rowspan="2">Overt codec<br>(re-transcoded)</td><td rowspan="2">S2, S4 / W3</td><td>For lossless TranSteg transcoding:<br>impossible to detect</td></tr>
<tr><td>For lossy TranSteg transcoding:<br>hard to detect<br>(to be verified in this study)</td></tr>
</table>

In this paper we focus on TranSteg detection for the worst-case scenario from the warden point of view. We assume that the warden is capable of inspecting the traffic only in single location (the most realistic assumption). Moreover, we exclude those cases where lossless compression was utilized – as stated above, in these situations the warden is helpless. That is why we focus on the case SLWC3, i.e., that only re-transcoded voice is available and a lossy pair of overt/covert codecs was used, i.e., scenario S4 and localization W3.

It must be emphasized that especially for this scenario TranSteg steganalysis is harder to perform than for most of the existing VoIP steganographic methods. This is because after the steganogram reaches the receiver, the hidden information is extracted and the speech data is practically restored to the originally sent data. As mentioned above, this is a huge advantage compared with existing VoIP steganographic methods, where the hidden data can be extracted and removed but the original data cannot be restored because it was previously erased due to a hidden data insertion process.

## 4. TranSteg Steganalysis Experimental Results

### 4.1 Experiment Methodology

As mentioned in the previous section, in our experiments we decided to check the possibility of TranSteg detection in the S4 scenario, when no reference signal is available, i.e., when a single warden is used at location W3 (case SLWC3). Since a comparison with the original data is not possible, we decided to use a detection method based on comparing parameters of the received signal against models of a normal (without TranSteg) and abnormal (with TranSteg) output speech signal.

We chose Mel-Frequency Cepstral Coefficients (MFCC) as the type of parameters to be extracted from the speech signal. The MFCC parameters have been successfully used in speech analysis since the 1970s and have been continuously employed in both speech and speaker recognition [7], as they have proved able to describe efficiently spectral features of speech. On the other hand, lossy speech codecs affect the speech spectrum, e.g., by smoothing the spectral envelope of the signal, so we hoped that the MFCC parameters would be helpful in detecting transcoding present in TranSteg. The same parameters have already been used in steganalysis in [18] (see Section 2), where they fed an SVM-based classifier.

In our approach, however, as a modeling method we decided to use Gaussian Mixture Models (GMMs) [26], since, combined with MFCCs, they have proved successful in various applications, including text-independent speaker recognition [30] and language recognition [27]. An expectation-maximization (EM) algorithm was used for GMM training. Fig. 3, created during one of experiments in this study, shows that MFCC parameters combined with GMM modeling are able to capture the differences between speech with and without TranSteg.

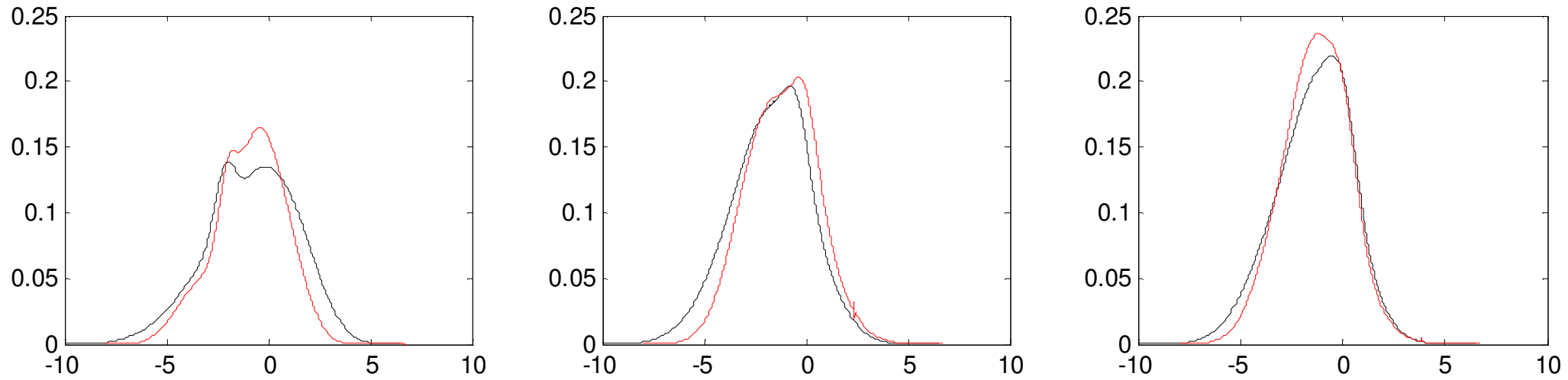

**Fig. 3:** Comparison of Gaussian mixture densities for normal G.711 transmission (black line) and transmission with TranSteg in S4 scenario (red line) for G.711/G.726 configuration. The first (left), second (middle) and third (right) MFCC coefficients are shown.

A series of experiments for various overt/covert pairs of codecs were conducted, including all the pairs which were recommended in [13], due to their achievable low steganographic cost and high steganographic bandwidth.

For each overt/covert codec pair, the experiment consisted of the following stages:

- A GMM model for **normal** speech transmission (no TranSteg) using a codec X was trained based on MFCC parameters extracted from the training speech signal;
- A GMM model for **abnormal** speech transmission (TranSteg active) using a pair of codecs X/Y was trained based on MFCC parameters extracted from the training speech signal;
- Using the two above GMM models, we checked if it is possible to recognize normal (no TranSteg) from abnormal (TranSteg active) transmission for a speech signal from test corpora.

Speech analysis was performed with an analysis window of 30 ms and analysis step of 10 ms. We used GMM models with 16 Gaussians and diagonal covariance matrixes. Transcoding was performed using the SoX package [22], Speex emulation [31] and “G.723.1 Speech Coder and Decoder” [15] library. Packet losses were not considered in this study. The number of MFCC parameters, as well as the length of testing signal, were subjects of experiments, the results of which will be presented in the next section.

Speech data used in experiments was extracted from five different speech corpora:

- TIMIT [9], containing speech data from 630 speakers of eight main dialects of US English, each of them uttering 10 sentences;
- TSP Speech corpus [16], containing 1400 recordings from 24 speakers, originally recorded with 48 kHz sampling, but also filtered and sub-sampled to different sample rates;
- CHAINS corpus [4], with 36 speakers of Hiberno-English recorded under a variety of speaking conditions;
- CORPORA – a speech database for Polish [10], containing over 16,000 recordings of 37 native Polish speakers reading 114 phonetically rich sentences and a collection of first names;
- AHUMADA – a spoken corpus for Castilian Spanish [24], containing recordings of 104 male voices, recorded in several sessions in various conditions (in situ and telephony speech, read and spontaneous speech, etc.).

GMM models for normal and abnormal transmissions were trained using 1600 recordings from the TIMIT corpus, originating from 200 speakers, each of them saying eight various sentences (two of the so-called SA TIMIT sentences were omitted because they were the same for all speakers, thus they could bias the acoustic model). In total, 90 minutes of speech were used to train both normal and abnormal models in each of the overt/covert scenarios.

Testing TranSteg detection was performed using the following test sets:

- 50 speakers from the TIMIT corpus, different from the ones used for training, hereinafter denoted as TIM;
- 23 speakers from the TSP Speech corpus from the “16k-LP7” subset, hereinafter denoted as TSP;

- 36 speakers from the CHAINS corpus from the “solo” subset, hereinafter denoted as CHA;
- 37 adult speakers from the CORPORA corpus, hereinafter denoted as COR;
- 25 male speakers from the AHUMADA corpus from in situ recordings (read speech), hereinafter denoted as AHU.

So the three first test corpora contained speech in English and the last two ones in Polish and Spanish, respectively. Each speech signal being tested contained recordings of one speaker only, to imitate the most common case if analyzing one channel of a VoIP conversation. Both training and testing were realized in the Matlab® environment using the h2m toolkit [23].

### 4.2 Experimental Results

The experiments were evaluated by calculating the recognition accuracy as the percentage of correct detections of normal and abnormal transmissions against all recognition trials. Results as low as around 50% mean recognition accuracy at a chance level; a result of 100% would mean an errorless detection of the presence or absence of TranSteg.

The first experiments were run to estimate the length of speech data required for effective steganalysis of TranSteg. Since the technique applied is based actually on statistical analysis of spectral parameters of speech, the amount of data required for analysis must be sufficiently high – such an analysis cannot be performed on speech extracted from a single 20 ms VoIP packet, or even from a few packets in a row. We ran our experiments on test signals ranging from 260 ms to 10 s; if we consider 20 ms packets, these correspond to the range between 13 and 500 voice packets.

The results of TranSteg recognition accuracy show that in some cases the accuracy grows steadily as the length of speech data increases, and becomes saturated after ca. 5-6 s (see the G.711/G.726 case presented in Fig. 4 on the left). It turns out that steganalysis based on a hundred 20-ms packets with speech data from the CHAINS corpus is successful with only 70% accuracy, but if we have a signal four times longer (8 s) the accuracy exceeds 90%. This means that in this case TranSteg needs to be active for a longer time in order to be spotted. In other cases (see e.g., the G.711/Speex7 pair in Fig. 4 on the right, or Speex7/iLBC), the recognition accuracy initially grows, but after 2-3 s it starts to oscillate around certain levels of accuracy. As an outcome of these experiments, for further analyses we decided to choose 7 s long speech signals.

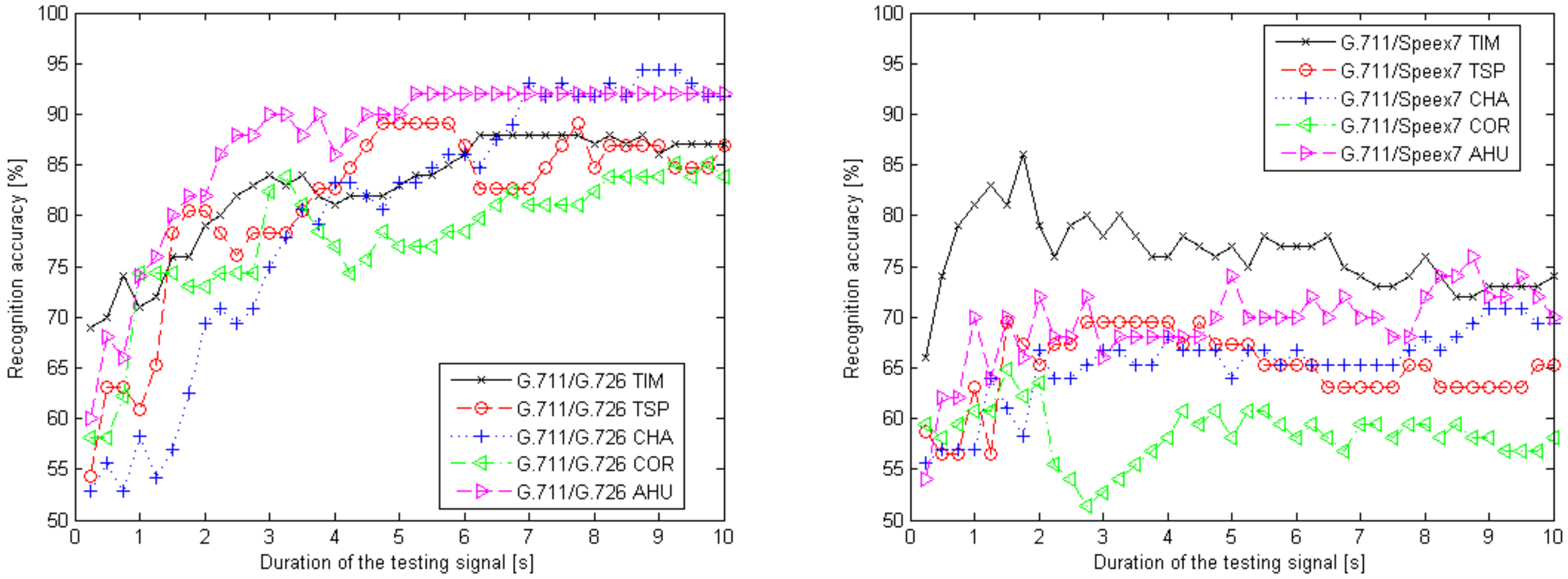

**Fig. 4:** TranSteg recognition accuracy vs. duration of the test signal, for G.711/G.726 (left) and G.711/Speex7 (right) configurations, for various test sets.

Next, experiments were aimed at deciding how many MFCC coefficients are needed for efficient TranSteg detection. In speech recognition usually 12 coefficients are used, usually with dynamic derivatives. In speaker recognition 12, 16, 19 or even 21 coefficients are used, in order to capture individual characteristics of a speaker ([3], [14]). Since here we are dealing with a different task, the number of MFCC coefficients required experimental verification. We checked the recognition accuracy for various overt/covert pairs of codecs for the number of MFCC coefficients ranging from 1 to 19.

The results show that in most cases the increase of the number of MFCC coefficients is beneficial, as presented for the configuration G.711/GSM06.10 in Fig. 5 (left). It is noteworthy that for the AHU, CHA and COR test sets, recognition with less than 5 MFCCs is very poor. On the other hand, in some cases, as shown in Fig. 5 (right) for iLBC/AMR, when the number of MFCCs exceeds 10-12 the recognition accuracy starts to decrease. As a conclusion it was decided to use 19 MFCC parameters in most cases, and 12 MFCC parameters for just a few cases: G.711/G.726, G.711/Speex7, G.711/AMR, iLBC/GSM 06.10 and iLBC/AMR.

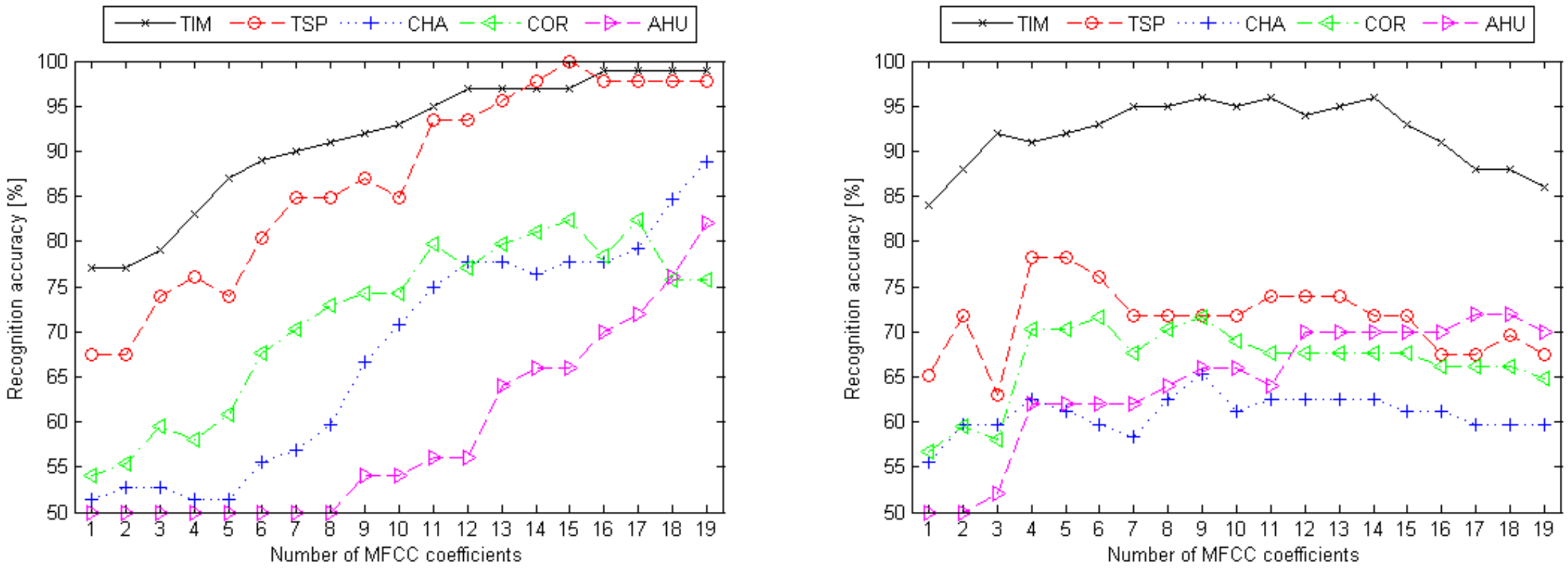

**Fig. 5:** TranSteg recognition accuracy vs. number of MFCC coefficients used in recognition, for G.711/GSM06.10 (left) and iLBC/AMR (right) configurations, for various test sets.

The detailed results of TranSteg recognition for various overt/covert codec configurations and various test sets are presented in Table II. It shows that the performance varies from slightly over 58% (which is close to random) for G.711/Speex7 for the COR test set, up to 100% for Speex7/G.729 for the TIM test set. In general, the results for TIM usually outperformed the remaining test sets. This is understandable considering the fact that other data from the same corpus (TIMIT) was used to train speech models, so similarities of recording conditions turned out to be an advantageous factor. This is why the results presented in Fig. 6 exclude the TIMIT corpus, and instead show the recognition results for the remaining data sets on average, as well as being divided into English and non-English data sets.

Table II. TranSteg recognition accuracy for various overt/covert configurations.

| Overt | Covert | TIM (EN) | TSP (EN) | CHA (EN) | COR (PL) | AHU (ES) |
|---|---|---|---|---|---|---|
| G711 | G.726 | 93.00 | 91.30 | 97.22 | 95.95 | 94.00 |
| | Speex7 | 80.00 | 67.39 | 59.72 | 58.11 | 68.00 |
| | iLBC | 95.00 | 97.83 | 90.28 | 63.51 | 80.00 |
| | GSM06.10 | 99.00 | 97.83 | 88.89 | 75.68 | 82.00 |
| | AMR | 92.00 | 82.61 | 87.50 | 82.43 | 74.00 |
| | G.729 | 99.00 | 97.83 | 91.67 | 86.49 | 82.00 |
| | G.723.1 | 99.00 | 91.30 | 72.22 | 90.54 | 74.00 |
| Speex7 | iLBC | 99.00 | 86.96 | 94.44 | 83.78 | 80.00 |
| | GSM06.10 | 97.00 | 95.65 | 79.17 | 79.73 | 68.00 |
| | AMR | 93.00 | 78.26 | 79.17 | 74.32 | 72.00 |
| | G.729 | 100.00 | 93.48 | 93.06 | 95.95 | 76.00 |
| | G.723.1 | 99.00 | 93.48 | 66.67 | 90.54 | 66.00 |
| iLBC | AMR | 96.00 | 73.91 | 62.50 | 67.57 | 64.00 |
| | GSM06.10 | 98.00 | 89.13 | 70.83 | 78.38 | 64.00 |
| | G.729 | 94.00 | 71.74 | 69.44 | 72.97 | 72.00 |
| | G.723.1 | 93.00 | 76.09 | 66.67 | 75.68 | 64.00 |

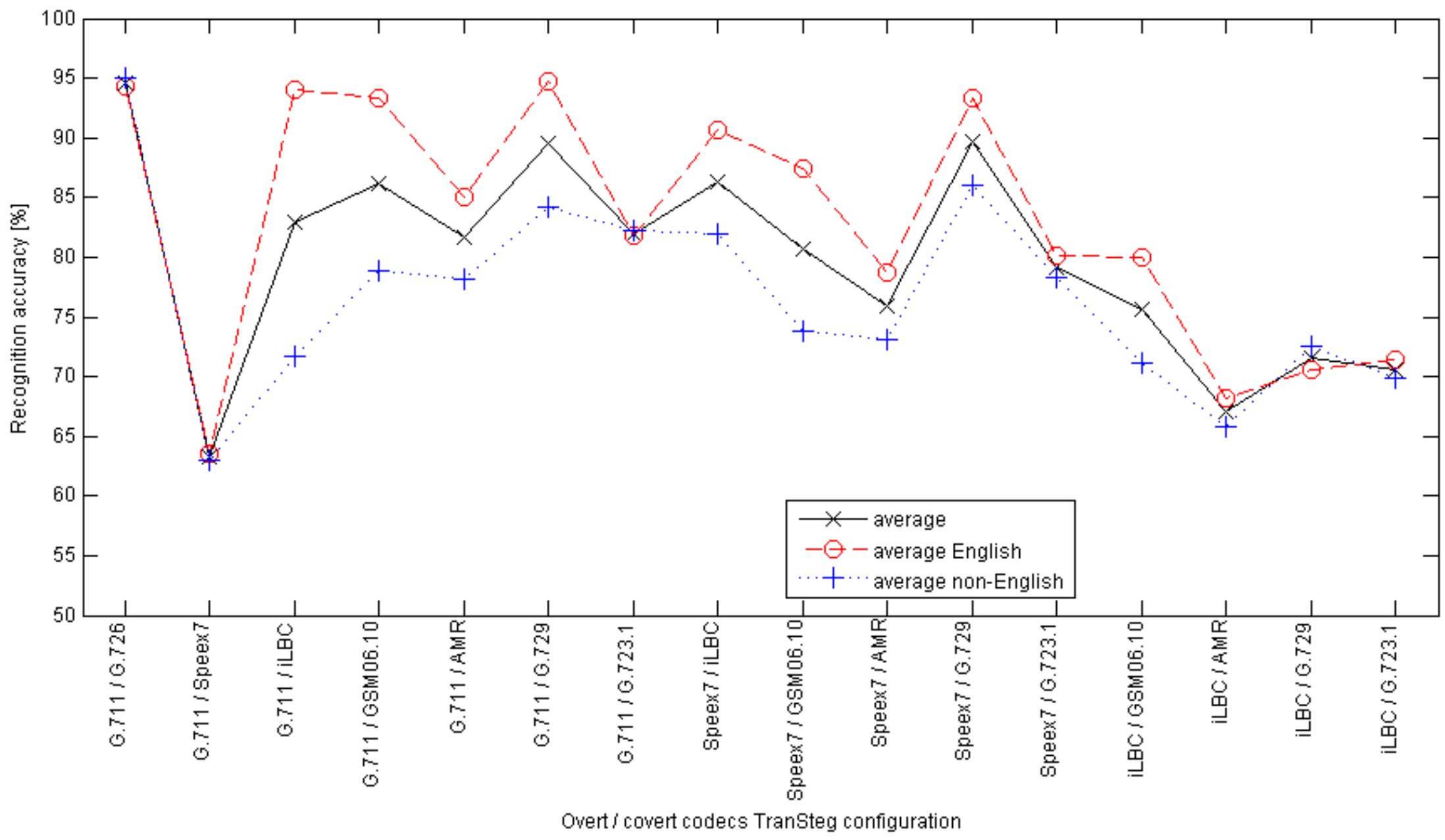

**Fig. 6:** Average TranSteg recognition accuracy for various overt/covert codec configurations, for English and non-English data sets (excluding TIM).

Both Table II and Fig. 6 show that pairs: G.711/Speex7, Speex7/AMR and the configurations with iLBC as the overt codec are quite resistant to steganalysis using the described method. The most resistant G.711/Speex7 and iLBC/AMR configurations can be detected with average recognition accuracy of only 63.3% and 67%, respectively. Other pairs with G.711 as the overt codec are much easier to detect (provided that we analyze enough speech data, in this case: 7 s), for example, the pair G.711/G.726 was detected with 94.6% accuracy. So was the pair Speex7/G.729, for which the presence (or absence) of TranSteg was correctly recognized in 90% of cases.

We found some correlation between steganographic cost and detectability of TranSteg: for example, the Speex7/G.729 pair offers a relatively high steganographic cost of 0.74 MOS, and at the same time it can be relatively easily detected (90% accuracy); the pair iLBC/AMR allows for TranSteg transmission with the cost of 0.46 MOS only, and is also difficult to detect. There are, however, a few exceptions to this rule: for example, the three covert codecs (G.726, AMR, Speex7) offering similar steganographic cost with G.711 to the overt one (ca. 0.4 MOS, see Fig. 2) behave quite differently as concerns the TranSteg detectability: G.711/G.726 can be recognized quite easily, whilst G.711/Speex7 proved to be the most resistant to steganalysis using the GMM/MFCC technique.

In general, TranSteg configurations with Speex7 and AMR as the covert codecs proved to be the most difficult to detect. This is confirmed in Fig. 7 (left). Fig. 6 and Fig. 7 show that the test sets for English were usually better recognized than non-English ones. This can be explained by the fact that the normal and abnormal speech models were trained just for English. Interestingly, a few configurations turned out to be "language-independent", e.g., the pairs with G.723.1 and Speex7 as the covert codec have the same TranSteg recognition accuracy results for both English and non-English data sets (see Fig. 6 and Fig. 7).

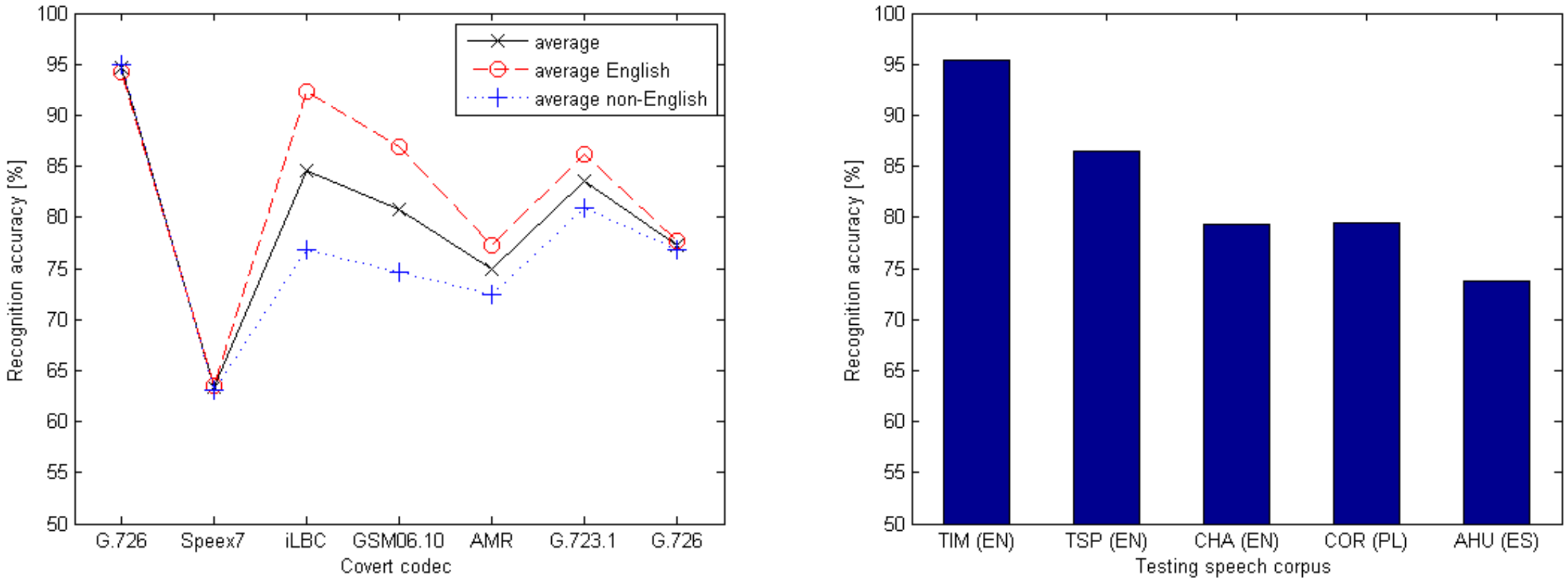

**Fig. 7:** Average TranSteg recognition accuracy for various covert codecs (left) and test sets (right).

## 5. Conclusions and Future Work

TranSteg is a fairly new steganographic method dedicated to multimedia services like IP telephony. In this paper the analysis of its detectability was presented for a variety of TranSteg scenarios and potential warden configurations. Particular attention was turned towards the very demanding case of a single warden located at the end of the VoIP channel (scenario S4). A steganalysis method based on the MFCC parameters and GMM models was described, implemented and thoroughly tested.

The results showed that the proposed method allowed for efficient detection of some codec pairs, e.g., G.711/G.726, with an average detection probability of 94.6%, or Speex7/G.729 with 89.6% detectability, or Speex7/iLBC, with 86.3% detectability. On the other hand, some TranSteg pairs remained resistant to detection using this method, e.g., the pair iLBC/AMR, with an average detection probability of 67%, which we consider to be low. This confirms that TranSteg with properly selected overt and covert codecs is an efficient steganographic method if analyzed with a single warden.

Successful detection of TranSteg using the described method, for a single warden at the end of the channel, requires at least 2 s of speech data to analyze, i.e., a hundred 20-ms VoIP packets. This should not be a problem, considering the fact that phone conversations last for minutes. However, if the overt channel contained not speech, but a piece of music, noise or just silence, the detectability of TranSteg would be seriously affected.

It must also be noted that, especially for the inspected hidden communication scenario (S4), TranSteg steganalysis is harder to perform than most of the existing VoIP steganographic methods. This is because, after the steganogram reaches the receiver, the hidden information is extracted and the speech data is practically restored to the data originally sent. If changes are made to the signal, they are not easily visible without a proper spectral and statistical analysis. This is a huge advantage compared with existing VoIP steganographic methods, where the hidden data can be extracted and removed but the original data cannot be restored because it was previously erased due to a hidden data insertion process.

Future work will include developing an effective steganalysis method when encryption using SRTP is utilized.

## ACKNOWLEDGMENTS

This research was partially supported by the Polish Ministry of Science and Higher Education and Polish National Science Centre under grants: 0349/IP2/2011/71 and 2011/01/D/ST7/05054.